\documentclass
[preprint,prd,onecolumn,showpacs,showkeys,nobibnotes,nofootinbib,titlepage]{revtex4}%
\usepackage{amsfonts}
\usepackage{amsmath}
\usepackage{amssymb}
\usepackage{graphicx}
\usepackage{float}
\usepackage{rotating}%
\providecommand{\U}[1]{\protect\rule{.1in}{.1in}}

\ifx\pdfoutput\relax\let\pdfoutput=\undefined\fi
\newcount\msipdfoutput
\ifx\pdfoutput\undefined\else
\ifcase\pdfoutput\else
\ifx\paperwidth\undefined\else
\ifdim\paperheight=0pt\relax\else\pdfpageheight\paperheight\fi
\ifdim\paperwidth=0pt\relax\else\pdfpagewidth\paperwidth\fi
\fi\fi\fi
\begin{document}
\preprint{ }
\title[Astrophysical Tests of Kinematical Conformal Cosmology]{Astrophysical Tests of Kinematical Conformal Cosmology in Fourth-Order
Conformal Weyl Gravity}
\author{Gabriele U. Varieschi}
\affiliation{Department of Physics, Loyola Marymount University - Los Angeles, CA 90045,
USA\footnote{Email: gvarieschi@lmu.edu}}

\begin{abstract}
In this work we analyze kinematical conformal cosmology (KCC), an alternative
cosmological model based on conformal Weyl gravity (CG), and test it against
current type Ia supernova (SNIa) luminosity data and other astrophysical observations.

Expanding upon previous work on the subject, we revise the analysis of SNIa
data, confirming that KCC can explain the evidence for an accelerating
expansion of the Universe without using dark energy or other exotic
components. We obtain an independent evaluation of the Hubble constant,
$H_{0}=67.53\
\operatorname{km}%
\
\operatorname{s}%
^{-1}\ \mathrm{Mpc}^{-1}$, very close to the current best estimates. The main
KCC and CG parameters are re-evaluated and their revised values are found to
be close to previous estimates.

We also show that available data for the Hubble parameter as a function of
redshift can be fitted using KCC and that this model does not suffer from any
apparent age problem. Overall, KCC remains a viable alternative cosmological
model, worthy of further investigation.

\end{abstract}
\startpage{1}
\endpage{ }
\maketitle
\tableofcontents

\section{\label{sect:introduction}Introduction}

Alternative theories of gravity (for reviews see \cite{Mannheim:2005bfa},
\cite{Clifton:2011jh}) have become more popular in recent years due to their
ability to account for astrophysical observations without using dark matter
(DM) and dark energy (DE). However, the cosmological constant - cold dark
matter model ($\Lambda$CDM) remains the standard explanation of current
astrophysical knowledge \cite{Beringer:1900zz}.

Fourth-order conformal Weyl gravity (CG, for short, in the following) is the
name given to an alternative gravitational theory, following the original work
by Weyl \cite{Weyl:1918aa}, not to be confused with other theories based on
conformal invariance. It was shown that CG\ (\cite{Mannheim:1988dj},
\cite{Kazanas:1988qa}) can describe the rotation curves of galaxies without DM
(\cite{Mannheim:2005bfa}, \cite{Mannheim:1992vj}, \cite{Mannheim:1996rv},
\cite{Mannheim:2010ti}, \cite{Mannheim:2010xw}, \cite{2012MNRAS.421.1273O},
\cite{Mannheim:2012qw}) and can give rise to the accelerated expansion of the
universe without resorting to DE (\cite{Mannheim:2005bfa},
\cite{2001ApJ...561....1M}).

A similar, but different approach to conformal cosmology was proposed by the
current author in a series of papers (\cite{Varieschi:2008fc},
\cite{Varieschi:2008va}, \cite{Varieschi:2010xs}) introducing a model which
was called kinematical conformal cosmology \cite{Varieschi:2008fc} (KCC in the
following) since it was based on purely kinematic considerations, without
using any dynamical equation of state for the Universe. This model was able to
account for the accelerated expansion of the Universe \cite{Varieschi:2008va}
and might also be able to explain the origin of some gravitational anomalies,
such as the Pioneer Anomaly \cite{Varieschi:2010xs} and the Flyby Anomaly
\cite{Varieschi:2014ata}.

Both models, the `standard' CG cosmology by Mannheim and KCC, were critically
analyzed by Diaferio et al. \cite{Diaferio:2011kc} and compared to standard
$\Lambda$CDM cosmology by applying a Bayesian approach to available
astrophysical data from type Ia supernovae (SNIa) and gamma-ray bursts.
Contrary to the authors' expectations \cite{Diaferio:2011kc}, the results of
this analysis showed that $\Lambda$CDM, Mannheim's CG, and KCC can all
describe the current astrophysical data equally well. Therefore, models based
on conformal gravity can be considered viable alternatives to $\Lambda$CDM and
are worthy of further investigation.

In addition, a recent study by Yang et al. \cite{Yang:2013skk} has tested
Mannheim's CG against recent astrophysical data from SNIa, determinations of
the Hubble parameter at different redshift, and in relation to the `age
problem' of the old quasar APM\ 08279+5255 at $z=3.91$. The outcome of this
analysis is that CG can describe all these astrophysical data in a
satisfactory manner and does not suffer from an age problem, as opposed to the
case of $\Lambda$CDM.

Following this recent work, the goal of this paper is to test our KCC against
the same astrophysical data used in Ref. \cite{Yang:2013skk} in order to
ascertain whether KCC is still a viable cosmological model. In Sect.
\ref{sect:conformal}, we begin by reviewing the main results of conformal
gravity and KCC. In Sect. \ref{sect:supernovae}, the main part of our paper,
we will constrain the KCC parameters, by using the latest Union 2.1 SNIa data,
and show that KCC can produce Hubble plots of the same quality as those
obtained with standard $\Lambda$CDM. In Sect. \ref{sect:hubble}, we will
compare the experimental data for the Hubble parameter, as a function of
redshift $z$, with KCC predictions and also briefly analyze the age problem in
the context of KCC.


\section{\label{sect:conformal}Conformal gravity and kinematical conformal
cosmology}

Conformal Gravity is based on the Weyl action:%

\begin{equation}
I_{W}=-\alpha_{g}\int d^{4}x\ (-g)^{1/2}\ C_{\lambda\mu\nu\kappa}%
\ C^{\lambda\mu\nu\kappa}, \label{eqn2.1}%
\end{equation}
where $g\equiv\det(g_{\mu\nu})$, $C_{\lambda\mu\nu\kappa}$ is the conformal or
Weyl tensor, and $\alpha_{g}$ is a dimensionless coupling constant. $I_{W} $
is the unique general coordinate scalar action that is invariant under local
conformal transformations: $g_{\mu\nu}(x)\rightarrow e^{2\alpha(x)}g_{\mu\nu
}(x)=\Omega^{2}(x)g_{\mu\nu}(x)$. CG does not suffer from the cosmological
constant problem and is renormalizable \cite{2012FoPh...42..388M}; it is a
ghost-free theory (\cite{2008PhRvL.100k0402B}, \cite{2008PhRvD..78b5022B}),
although it still faces some theoretical challenges
(\cite{2006PhRvD..74b3002F}, \cite{2004CQGra..21.1897P},
\cite{2006MNRAS.369.1667H}, \cite{2009MNRAS.393..215D},
\cite{1994ApJ...428...17E}, \cite{Yoon:2013rxa}).

The fourth-order CG field equations, $4\alpha_{g}W_{\mu\nu}=T_{\mu\nu}$ (where
$W_{\mu\nu}$ is the Bach tensor---see \cite{Mannheim:2005bfa},
\cite{Varieschi:2008fc} for full details), were studied in 1984 by Riegert
\cite{Riegert:1984zz}, who obtained the most general, spherically symmetric,
static electrovacuum solution. The explicit form of this solution, for the
practical case of a static, spherically symmetric source in CG, i.e., the
fourth-order analogue of the Schwarzschild exterior solution in General
Relativity (GR), was then derived by Mannheim and Kazanas in 1989
(\cite{Mannheim:1988dj}, \cite{Kazanas:1988qa}). This latter solution, in the
case $T_{\mu\nu}=0$ (exterior solution), is described by the metric%
\begin{equation}
ds^{2}=-B(r)\ c^{2}dt^{2}+\frac{dr^{2}}{B(r)}+r^{2}(d\theta^{2}+\sin^{2}%
\theta\ d\phi^{2}), \label{eqn2.2}%
\end{equation}
with%
\begin{equation}
B(r)=1-3\beta\gamma-\frac{\beta(2-3\beta\gamma)}{r}+\gamma r-\kappa r^{2}.
\label{eqn2.3}%
\end{equation}

The three integration constants in the last equation\ are as follows:
$\beta\ (%
\operatorname{cm}%
)$ can be considered the CG\ equivalent of the geometrized mass $\frac
{GM}{c^{2}}$, where $M$ is the mass of the (spherically symmetric) source and
$G$ is the universal gravitational constant; two additional parameters,
$\gamma\ (%
\operatorname{cm}%
^{-1})$ and $\kappa\ (%
\operatorname{cm}%
^{-2})$, are required by CG, while the standard Schwarzschild solution is
recovered for $\gamma,\kappa\rightarrow0$ in the equations above. The
quadratic term $-\kappa r^{2}$ indicates a background De Sitter spacetime,
which is important only over cosmological distances, since $\kappa$ has a very
small value. Similarly, $\gamma$ measures the departure from the Schwarzschild
metric at smaller distances, since the $\gamma r$ term becomes significant
over galactic distance scales.

The values of the CG parameters were first determined by Mannheim
\cite{Mannheim:2005bfa}:\footnote{Other estimates of these parameters exist in
the literature. For example, in Ref. \cite{Sultana:2012qp}, constraints on the
value of the $\gamma$ constant were obtained by studying the perihelion shift
of planetary motion in CG.}%
\begin{equation}
\gamma=3.06\times10^{-30}%
\operatorname{cm}%
^{-1},\ \kappa=9.54\times10^{-54}%
\operatorname{cm}%
^{-2}. \label{eqn2.4}%
\end{equation}
In our previous KCC publications (\cite{Varieschi:2008fc},
\cite{Varieschi:2008va}) we have shown a different way to compute the
CG\ parameters, obtaining values which differ by a few orders of magnitude
from those above:%
\begin{equation}
\gamma=1.94\times10^{-28}%
\operatorname{cm}%
^{-1},\ \kappa=6.42\times10^{-48}%
\operatorname{cm}%
^{-2}. \label{eqn2.5}%
\end{equation}
We will revise and update the values of these parameters in Sect.
\ref{sect:supernovae} by constraining them with recent astrophysical data.

Mannheim et al. (\cite{Mannheim:2005bfa}, \cite{Mannheim:1992vj},
\cite{Mannheim:1996rv}, \cite{Mannheim:2010ti}, \cite{Mannheim:2010xw},
\cite{2012MNRAS.421.1273O}, \cite{Mannheim:2012qw}) used the CG solutions in
Eqs. (\ref{eqn2.2})-(\ref{eqn2.3}) to perform extensive data fitting of
galactic rotation curves without any DM contribution, with the values of
$\gamma$ and $\kappa$ as in Eq. (\ref{eqn2.4}). Although the values of these
CG\ parameters are very small, the linear and quadratic terms in Eq.
(\ref{eqn2.3}) become significant over galactic and/or cosmological distances.

This also means that CG solutions (including those for other types of sources,
see discussion in \cite{Varieschi:2014ata}) are not asymptotically flat, thus
raising the question of possible `gravitational redshift' effects at large
distances. In fact, this was the main motivation for our `kinematical
approach' to conformal cosmology: in regions far away from massive sources
(for $r\gg\beta(2-3\beta\gamma)$) and also ignoring the term $\beta\gamma$, as
suggested by the analysis of galactic rotation velocities, $B(r)$ simplifies to%

\begin{equation}
B(r)=1+\gamma r-\kappa r^{2}. \label{eqn2.6}%
\end{equation}

This implies a possible gravitational redshift at large distances, analogous
to the one experimentally observed in standard GR near massive sources such as
the Earth, the Sun, or white dwarfs. This effect is related to the square-root
of the ratio of the time-time components $g_{00}$ of the metric at two
different locations. In Ref. \cite{Varieschi:2008fc} we considered our current
spacetime location ($r=0$; $t_{0}$) in relation to the spacetime location
($r>0$; $t<t_{0}$) of a distant galaxy which emits light at a time $t$ in the
past that reaches us at present time $t_{0}$ and appears to be redshifted in
relation to the standard redshift parameter $z$.

We then argued that this observed redshift could be due (in part, or totally)
to the gravitational redshift effect mentioned above. If this effect were
indeed the only source of the observed redshift, with the metric in Eq.
(\ref{eqn2.6}), we would have:%

\begin{equation}
1+z=\sqrt{\frac{-g_{00}(0,t_{0})}{-g_{00}(r,t)}}=\frac{1}{\sqrt{1+\gamma
r-\kappa r^{2}}}. \label{eqn2.7}%
\end{equation}
In other words, if the CG metric in Eqs. (\ref{eqn2.2})-(\ref{eqn2.3}) has a
true physical meaning, as it seems to be the case from the detailed fitting of
galactic rotational curves, it should also determine strong gravitational
redshift at very large cosmological distances.\footnote{Eq. (\ref{eqn2.6}) is
valid for regions far away from massive sources, i.e., for $r\gg\beta
(2-3\beta\gamma)\simeq2\beta\simeq2\frac{GM}{c^{2}}$, where $M$ can be
considered the mass of the largest structures in our Universe, such as
galaxies, or clusters of galaxies. Therefore, the resulting characteristic
distance $r$ represents the scale at which our kinematical approach is
appropriate. For example, considering the estimated mass of a cluster, or a
supercluster of galaxies, the resulting characteristic distance is
approximately $r\gtrsim0.1-10\ Mpc$, which shows that KCC mainly applies to
the inter-galactic or cosmological scale.} As far as we are aware, this issue
has never been raised in all current CG literature (except, of course, in our
previous papers).

The CG metric in Eqs. (\ref{eqn2.2}) and (\ref{eqn2.6}) is actually conformal
to the standard FRW metric (see details in \cite{Mannheim:1988dj}\ or
\cite{Varieschi:2008fc}):%

\begin{equation}
ds^{2}=-c^{2}d\mathbf{t}^{2}+\mathbf{a}^{2}(\mathbf{t})\left[  \frac
{d\mathbf{r}^{2}}{1-\mathbf{kr}^{2}}+\mathbf{r}^{2}(d\theta^{2}+\sin^{2}%
\theta\ d\phi^{2})\right]  , \label{eqn2.8}%
\end{equation}
where $\mathbf{a}(\mathbf{t})$ is the standard Robertson-Walker scale factor,
$\mathbf{k}=k/\left\vert k\right\vert =0,\pm1$ and $k=-\gamma^{2}/4-\kappa$.
As in our previous papers, we distinguish here between two sets of
coordinates: the Static Standard Coordinates - SSC $(r,t) $ used in Eqs.
(\ref{eqn2.2})-(\ref{eqn2.3}) and (\ref{eqn2.6})-(\ref{eqn2.7}), as opposed to
the FRW coordinates $(\mathbf{r,t})$ ---in bold--- used in Eq. (\ref{eqn2.8}%
).\footnote{Similarly, bold type characters will be used for quantities
referring to the FRW geometry, while normal type characters will be used with
reference to the SSC coordinates. For example, the RW scale factor will be
denoted here as $\mathbf{a(t)}$ or $a(t)$, respectively, in the two cases. In
our previous papers we used $\mathbf{R(t)}$ and $R(t)$ for the scale factor,
but we now prefer to adopt the more common notation, $\mathbf{a(t)}$ or
$a(t)$, in this work.} Full details of the complete transformations between
these coordinates can be found in our Refs. \cite{Varieschi:2008fc} and
\cite{Varieschi:2008va}.

This local conformal invariance induces a dependence of the length and time
units on the local metric, so that the observed redshift can be interpreted as
the ratio between the wavelength $\lambda(\mathbf{r,t})$ of the radiation
emitted by atomic transitions, at the time and location of the source, and the
wavelength $\lambda(\mathbf{0,t}_{0})$ of the same atomic transition measured
here on Earth at current time. Since modern metrology defines our common units
of length $\delta l$ and time $\delta t$\ as being proportional respectively
to the wavelength and to the period (inverse of the frequency $\nu$) of
radiation emitted during certain atomic transitions, we can write the
following `redshift equation'%

\begin{equation}
1+z=\frac{\mathbf{a(t}_{0}\mathbf{)}}{\mathbf{a(t)}}=\frac{\lambda
(\mathbf{r,t})}{\lambda(\mathbf{0,t}_{0})}=\frac{\delta l(\mathbf{r,t}%
)}{\delta l(\mathbf{0,t}_{0})}=\frac{\nu(\mathbf{0,t}_{0})}{\nu(\mathbf{r,t}%
)}=\frac{\delta t(\mathbf{r,t})}{\delta t(\mathbf{0,t}_{0})}, \label{eqn2.9}%
\end{equation}
connecting wavelengths $\lambda$ to unit-lengths $\delta l$ and frequencies
$\nu$ to unit-time intervals $\delta t$ (we also use $\lambda\nu=c$, with a
constant speed of light $c$).

Therefore, in KCC the observed redshift is due to the change of length and
time units over cosmological spacetime, as opposed to the standard explanation
of a pure expansion of the scale factor $\mathbf{a}$. In view of this
interpretation, and connecting together Eqs. (\ref{eqn2.7}) and (\ref{eqn2.9}%
), KCC is able to derive directly the scale factor as a function of space or
time coordinates, without solving the dynamical field equations. In terms of
SSC, we have:%

\begin{equation}
1+z=\frac{a(0)}{a(r)}=\frac{1}{\sqrt{1+\gamma r-\kappa r^{2}}},
\label{eqn2.10}%
\end{equation}
or, using appropriate coordinate transformations, in terms of FRW coordinates:%

\begin{equation}
1+z=\frac{\mathbf{a(0)}}{\mathbf{a(r)}}=\sqrt{1-\mathbf{k\ r}^{2}}%
-\delta\mathbf{r}, \label{eqn2.11}%
\end{equation}
with%

\begin{equation}
\delta=\frac{\gamma}{2}%
\begin{Bmatrix}
\left\vert k\right\vert ^{-1/2} & for\ k\neq0\\
1 & for\ k=0
\end{Bmatrix}
. \label{eqn2.12}%
\end{equation}

All these scale-factor equations can also be written explicitly in terms of
the time coordinates $t$ and $\mathbf{t}$, as is usually done in standard
cosmology, by computing the time it takes for a light signal, emitted at
radial distance $r$ or $\mathbf{r}$, to reach the observer at the origin. The
detailed expressions for $a(t)$ and $\mathbf{a(t)}$, as well as all the
connecting formulas between the different variables and conformal parameters,
can be found in Ref. \cite{Varieschi:2008fc} (see Table I). Furthermore, from
the plots of the KCC scale factors, such as $\mathbf{a(r)}$ from Eq.
(\ref{eqn2.11}), it can be seen that the observed redshift $z>0$ is only
possible for $\mathbf{k}=-1$, so that the other two cases, $\mathbf{k}=0,+1$,
are actually ruled out.

The new CG dimensionless\footnote{The parameter $\delta$ in Eq. (\ref{eqn2.12}%
) is dimensionless only for $\mathbf{k}=\pm1$. For the $\mathbf{k}=0$ case,
Eq. (\ref{eqn2.11}) simply becomes $1+z=\mathbf{a(0)/a(r)}=1-\frac{\gamma}%
{2}\mathbf{r}$. In this particular case, the coordinate $\mathbf{r}$ has
dimensions of length, so this equation is still dimensionally correct (in this
case the scale factor $\mathbf{a}(\mathbf{t})$ becomes a dimensionless
quantity, so that Eq. (\ref{eqn2.8}) is also correct). This is due to the
particular form of the transformation between SSC and FRW coordinates, for the
special $k=0$ case. See Sect. 3.1 in Ref. \cite{Varieschi:2008fc} for complete
details.} parameter $\delta=\frac{\gamma}{2\sqrt{\left\vert k\right\vert }}$
(for $\mathbf{k}=-1$) in Eq. (\ref{eqn2.12}) becomes the most important
quantity in KCC: it combines together the original CG parameters $\gamma$ and
$\kappa$, in view also of the relation between $k$ and $\kappa$%

\begin{equation}
k=-\frac{\gamma^{2}}{4}-\kappa\label{eqn2.13}%
\end{equation}
already mentioned above.\footnote{In our previous papers, we considered the
possibility that all these CG parameters might also be changing with spacetime
coordinates. In particular, we supposed that the $\delta$ parameter might play
the role of a universal time and we used the zero subscript to denote the
current values of all these parameters (i.e., $\delta_{0}$, $\gamma_{0}$,
etc.). In this paper, we are just considering the current values of these
parameters, so we simply write $\delta$, $\gamma$, $\kappa$, etc.} It can be
shown that $\left\vert \delta\right\vert <1$ and that, for $\mathbf{k}=-1$,
Eq. (\ref{eqn2.11}) yields the following direct relation between $\mathbf{r}$
and $z$:%

\begin{equation}
\mathbf{r}=\frac{\delta(1+z)\pm\sqrt{(1+z)^{2}-(1-\delta^{2})}}{1-\delta^{2}}.
\label{eqn2.14}%
\end{equation}

The plus-minus sign in the last equation indicates that there are two
locations where $z=0$: at the origin $\mathbf{r}=0$, and at a particular
radial location $\mathbf{r}_{rs}=\frac{2\delta}{1-\delta^{2}}$ which becomes
of physical significance for $\delta>0$. In fact, in this particular case,
there is a region of negative redshift (i.e., a blueshift) for $0<\mathbf{r}%
<\mathbf{r}_{rs}$, followed by a standard redshift region at larger radial
distances, for $\mathbf{r}>\mathbf{r}_{rs}=\frac{2\delta}{1-\delta^{2}}$. This
suggests that the (current) value of $\delta$ should be small and positive, so
that the supposed blueshift region would be a small (practically undetectable)
region around the observer: for example, a small region of the size of the
Solar System, or similar.

In two of our previous papers (\cite{Varieschi:2008va},
\cite{Varieschi:2010xs}) we actually suggested that this local blueshift
region could have been the origin of the Pioneer Anomaly (PA - for a review,
see \cite{Turyshev:2010yf}) since `blueshifted' signals coming from the
Pioneer spacecraft would appear to be equivalent to the observed anomalous
acceleration. In view of this possible connection, the value of the $\gamma$
parameter in Eq. (\ref{eqn2.5}) was directly inferred from the Pioneer
anomalous acceleration (\cite{Varieschi:2008va}, \cite{Varieschi:2010xs}); the
value of the $\delta$ parameter was then computed \cite{Varieschi:2008va} from
the fitting of the SNIa data available at the time, and the values of the
parameters $k$ and $\kappa$ were obtained through Eqs. (\ref{eqn2.12}) and
(\ref{eqn2.13}). In summary, the values of the CG parameters were determined
as follows (see also Table 1 in Ref. \cite{Varieschi:2008va}):%

\begin{equation}
\delta=3.83\times10^{-5},\ \gamma=1.94\times10^{-28}%
\operatorname{cm}%
^{-1},\ k=-6.42\times10^{-48}%
\operatorname{cm}%
^{-2},\ \kappa=6.42\times10^{-48}%
\operatorname{cm}%
^{-2}. \label{eqn2.15}%
\end{equation}

Although it is still possible that the PA might have a gravitational origin,
i.e., due to modifications of GR, it is now widely accepted that the cause of
this anomaly is probably more mundane \cite{Turyshev:2012mc}: thermal recoil
forces originating from the spacecraft radioactive thermoelectric generators.
Therefore, in the following sections we will perform a new computation of the
CG parameters in Eq. (\ref{eqn2.15}), without using any more data related to
the PA. We will begin, in the following section, by constraining our
parameters using updated SNIa data.

\section{\label{sect:supernovae}KCC and type Ia supernovae}

In order to constrain the CG parameters with recent SNIa data we need to
redefine the luminosity distance in KCC, since this is the main cosmological
distance used in this context. In this section we will expand upon concepts
already introduced in Ref. \cite{Varieschi:2008va} (more details about the
definitions of distances in KCC can be found in this reference). We start by
noticing that the new interpretation of the redshift discussed in the previous
section (in particular, in Eq. (\ref{eqn2.9})) implies that lengths and time
intervals scale with redshift $z$ as:%

\begin{align}
\Delta l_{z}  &  =(1+z)\ \Delta l_{0}\label{eqn3.1}\\
\Delta t_{z}  &  =(1+z)\ \Delta t_{0},\nonumber
\end{align}
where the subscript $0$ indicates intervals of the given quantity associated
with objects which share the same spacetime location of the observer at the
origin (namely, here on Earth at $r=0$ and at our current time $t_{0}$), while
the subscript $z$ indicates intervals of the same quantity associated with
objects at redshift $z\neq0$, as seen or measured by the same observer at the origin.

It should be emphasized that this change in lengths, or time intervals (as
well as wavelengths, frequencies, and all other kinematical quantities derived
from lengths and times), is due to the spacetime location of the object being
studied (as measured by the redshift parameter $z$) and not to the `cosmic
expansion' as in the standard cosmological model.

It is natural to assume that masses, energies, luminosities, and other
dynamical quantities will follow similar scaling laws, but not necessarily the
same as the one in Eq. (\ref{eqn3.1}). In Ref. \cite{Varieschi:2008va} we
assumed the following scaling laws for masses and energies:\footnote{Mass and
energy will scale in the same way, since $\Delta E\propto\Delta l^{2}\Delta
t^{-2}\Delta m$, with lengths and times scaling in the same manner, due to Eq.
(\ref{eqn3.1}).}%

\begin{align}
\Delta m_{z}  &  =f(1+z)\ \Delta m_{0}\label{eqn3.2}\\
\Delta E_{z}  &  =f(1+z)\ \Delta E_{0},\nonumber
\end{align}
where $f(1+z)$ is some arbitrary function of $(1+z)$, so that $\lim
_{z\rightarrow0}f(1+z)=1$.

As a consequence of these scaling laws, the `absolute luminosity' $L$, or
energy emitted per unit time, will scale as%

\begin{equation}
L_{z}=\frac{f(1+z)}{\left(  1+z\right)  }\ L_{0}, \label{eqn3.3}%
\end{equation}
where the meaning of the subscripts is the same as described above for the
other quantities. Thus, KCC postulates a change in the absolute luminosity of
a `standard candle,' which is intrinsically due to its spacetime location,
while standard cosmology assumes an invariable absolute luminosity $L$ of the
standard candle being considered.

Standard cosmology defines the luminosity distance as $d_{L}=\sqrt{\frac
{L}{4\pi l}}=\mathbf{a}_{0}\mathbf{r}(1+z)$, with $L$ and $l$ being the
absolute and apparent luminosities of the standard candle being used as a
distance indicator; $\mathbf{a}_{0}$ denotes the current value of the scale
factor and the $(1+z)$ factor on the right-hand side of the equation
originates from a $(1+z)^{2}$ dimming factor under the square root. This
factor is due to the standard redshift of the photon frequency and also to a
time dilation effect of the emission interval of photons.

KCC considers instead this $(1+z)^{2}$ dimming factor as unphysical, so the
$(1+z)$ factor on the right-hand side of the standard luminosity distance
equation is completely eliminated. In view also of our scaling law for
luminosities in Eq. (\ref{eqn3.3}), and of Eq. (\ref{eqn2.14}), we then define
the luminosity distance in KCC as:\footnote{In the following equation we
choose the positive sign in front of the square root to select the solution
corresponding to past redshift, $z>0$ for $\mathbf{r}>\mathbf{r}_{rs}%
=2\delta/(1-\delta^{2})$, which is the correct choice for the following
analysis of SNIa data.}%

\begin{equation}
d_{L}\equiv\sqrt{\frac{L_{z}}{4\pi l}}=\sqrt{\frac{f(1+z)}{(1+z)}\frac{L_{0}%
}{4\pi l}}=\mathbf{a}_{0}\mathbf{r}=\mathbf{a}_{0}\frac{\delta(1+z)+\sqrt
{(1+z)^{2}-(1-\delta^{2})}}{(1-\delta^{2})}. \label{eqn3.4}%
\end{equation}
Since this definition assumes an intrinsic dimming of the luminosity $L_{z}$
with redshift $z$, it leads to distance estimates which are dramatically
different from those of standard cosmology for different values of $z$ (see
the first three columns in Table 2 of Ref \cite{Varieschi:2008va}).

To avoid this issue, an alternative definition could be employed, which would
retain the concept of an invariable luminosity $L_{0}$ of a standard candle,
while including the other aspects of KCC. We can obtain this alternative
luminosity distance $\widetilde{d}_{L}$ by modifying the previous equation as follows:%

\begin{equation}
\widetilde{d}_{L}\equiv\sqrt{\frac{L_{0}}{4\pi l}}=\sqrt{\frac{(1+z)}{f(1+z)}%
}\mathbf{a}_{0}\mathbf{r}=\sqrt{\frac{(1+z)}{f(1+z)}}\mathbf{a}_{0}%
\frac{\delta(1+z)+\sqrt{(1+z)^{2}-(1-\delta^{2})}}{(1-\delta^{2})},
\label{eqn3.5}%
\end{equation}
so that the right-hand side of the equation now depends explicitly on the
still unknown function $f(1+z)$. In Table 2 of Ref. \cite{Varieschi:2008va},
it was shown that distances estimated using $\widetilde{d}_{L}$ are very close
to those of standard cosmology (compare the values in the fourth column of
this table with those in the third or fifth columns), so the KCC definition in
Eq. (\ref{eqn3.5}) more closely agrees with the luminosity distance of
standard cosmology.

We will see in the following that both definitions, in Eqs. (\ref{eqn3.4}) and
(\ref{eqn3.5}), lead to the same results when applied to SNIa data, but they
differ conceptually: the former assumes a variable absolute luminosity $L_{z}$
of a standard candle, while the latter assumes an invariable absolute
luminosity $L_{0}$, which is more in line with the standard interpretation.

Before we can apply these definitions to the analysis of SNIa data, we need to
obtain an explicit form for the $f(1+z)$ function, which enters most of the
KCC equations above. Expanding upon the arguments discussed in our previous
work \cite{Varieschi:2008va}, we can assume the following properties for this function:

\begin{enumerate}
\item $f$ is some arbitrary function of $(1+z)$, with a `fixed point'\ at $1$,
that is, $f(1)=1$, or $\lim_{z\rightarrow0}\ f(1+z)=1$.

\item $f$ is a dimensionless quantity, so that Eqs. (\ref{eqn3.2}%
)-(\ref{eqn3.5}) are dimensionally correct.

\item $f$ is a function possibly built out of other expressions of KCC, which
also depend on the factor $(1+z)$.
\end{enumerate}

Although the last property in the list above is just an educated guess, it
suggests that the function $f$ might depend on the following KCC factor:%

\begin{equation}
\frac{d_{L}}{d_{REF}}=\frac{\delta(1+z)+\sqrt{(1+z)^{2}-(1-\delta^{2})}%
}{2\delta}, \label{eqn3.6}%
\end{equation}
constructed as the (dimensionless) ratio between the luminosity distance in
Eq. (\ref{eqn3.4}) and the reference distance%

\begin{equation}
d_{REF}=\mathbf{a}_{0}\mathbf{r}_{rs}=\mathbf{a}_{0}\frac{2\delta}%
{1-\delta^{2}}, \label{eqn3.7}%
\end{equation}
which corresponds to the value $\mathbf{r}_{rs}$ of the radial coordinate
(other than the origin) where we have $z=0$ (see discussion after Eq.
(\ref{eqn2.14})). Therefore, as it was argued also in Ref.
\cite{Varieschi:2008va}, $d_{REF}$ represents the ideal reference distance at
which we should place a `standard candle' of given absolute luminosity $L_{0}%
$: at this location its luminosity is not affected by the scaling effect of
Eq. (\ref{eqn3.3}), since $z=0$ for $\mathbf{r}=\mathbf{r}_{rs}$. In KCC
$d_{REF}$ is the equivalent of the standard reference distance of $10$
\textrm{parsec}, used for standard candles, such as supernovae.

Following the discussion above, the most general form of the function $f(1+z)
$ that we will consider is:%

\begin{equation}
f(1+z)=\frac{(1+z)^{\beta}}{\left(  \frac{d_{L}}{d_{REF}}\right)  ^{\alpha}%
}=\left[  \frac{2\delta}{\delta(1+z)+\sqrt{(1+z)^{2}-(1-\delta^{2})}}\right]
^{\alpha}(1+z)^{\beta}, \label{eqn3.8}%
\end{equation}
where $\alpha$ and $\beta$ are coefficients to be determined from SNIa data
fitting. Again, the choice of the function $f(1+z)$ in the previous equation
is just an educated guess, an `ansatz' based on the only two functions of
$(1+z)$ introduced in KCC:\ a function $(1+z)^{\beta}$, which generalizes the
simple $(1+z)$ scaling factor in Eq. (\ref{eqn3.1}), and a function $1/\left(
d_{L}/d_{REF}\right)  ^{\alpha}$, which generalizes the inverse-square
dependence of the apparent luminosity of a radiation source upon the
(luminosity) distance between the observer and the source.

\subsection{\label{sect:Union}SNIa data fitting}

In our previous work, we determined the CG parameters by using the SNIa data
available at the time (292 SNIa data of the `gold-silver' set, see
\cite{Varieschi:2008va}\ for details) and by considering the value of the
Pioneer anomalous acceleration. As already mentioned, we will not use the PA
data in this study, but we will use the latest compilation of SNIa data: the
580 supernovae from the Union 2.1 data set (\cite{Kowalski:2008ez},
\cite{2010ApJ...716..712A}, \cite{Suzuki:2011hu}).

The distance modulus $\mu$ (difference between the apparent magnitude $m$ and
the absolute magnitude $M$) is usually computed, using Pogson's law, in terms
of the logarithm of the ratio between the apparent luminosity $l_{z}$ (at
redshift $z$) and the reference apparent luminosity $l_{REF}$ (at the
reference distance of choice). It can then be expressed in terms of absolute
luminosities and distances, using the general relation $l=\frac{L}{4\pi
d_{L}^{2}}$. We have:%

\begin{equation}
\mu(z)=m(z)-M=-2.5\log_{10}\left(  \frac{l_{z}}{l_{REF}}\right)
=-2.5\log_{10}\left(  \frac{L_{z}}{L_{REF}}\frac{d_{REF}^{2}}{d_{L}^{2}%
}\right)  , \label{eqn3.9}%
\end{equation}
where the subscript $z$ refers to quantities evaluated at redshift $z\neq0$,
while the subscript $REF$ indicates the `reference' value of the quantity,
i.e., when the standard candle is placed at the reference distance.

As explained before, we have two possible choices for this reference distance:
the traditional distance of $10\ \mathrm{pc}$ (since usually the absolute
luminosity $L$ of a `standard candle' is defined as the apparent luminosity of
the same object placed at $10\ $\textrm{parsec}) and the KCC reference
distance $d_{REF}$ in Eq. (\ref{eqn3.7}) above, since this is the only
location, other than the origin, where $z=0$.

Using this latter choice for the reference distance and combining Eq.
(\ref{eqn3.9}) with Eqs. (\ref{eqn3.3}), (\ref{eqn3.4}), (\ref{eqn3.6}), and
(\ref{eqn3.8}), we obtain explicitly:%

\begin{align}
\mu(z)  &  =2.5(2+\alpha)\log_{10}(d_{L}/d_{REF})+2.5(1-\beta)\log
_{10}(1+z)\label{eqn3.10}\\
&  =2.5(2+\alpha)\log_{10}\left[  \frac{\delta(1+z)+\sqrt{(1+z)^{2}%
-(1-\delta^{2})}}{2\delta}\right]  +2.5(1-\beta)\log_{10}(1+z),\nonumber
\end{align}
an expression which can be used directly to fit SNIa data and determine the
value of the three free parameters $\alpha$, $\beta$, and $\delta$.

Using this last equation as a fitting formula for the Union 2.1 SNIa data, we
obtained the following `best-fit' values for the free parameters:%

\begin{equation}
\alpha=2.096\pm0.027,\ \beta=1.141\pm0.091,\ \delta=\left(  4.120\pm
0.221\right)  \times10^{-5}. \label{eqn3.11}%
\end{equation}
Assuming that $\alpha$ and $\beta$ are likely to be integer numbers, due to
their role in the definition of the function $f(1+z)$ in Eq. (\ref{eqn3.8}),
and close to the values reported in the previous equation, we repeated the
fitting procedure, first by setting $\beta=1$:%

\begin{equation}
\alpha=2.058\pm0.010,\ \beta=1,\ \delta=\left(  3.817\pm0.087\right)
\times10^{-5}, \label{eqn3.12}%
\end{equation}
then by fixing both $\alpha$ and $\beta$ as follows:%

\begin{equation}
\alpha=2,\ \beta=1,\ \delta=\left(  3.356\pm0.005\right)  \times10^{-5}.
\label{eqn3.13}%
\end{equation}

All these fits have good statistical quality ($R^{2}=0.996$) and clearly
confirm the results of our past SNIa data fitting \cite{Varieschi:2008va},
where it was a priori postulated that $\alpha=2$,$\ \beta=1$, and $\delta$ was
found to be as in Eq. (\ref{eqn2.15}). It can also be shown that our fitting
formula in the second line of Eq. (\ref{eqn3.10}) can even be obtained by
using the alternative definition of the luminosity distance $\widetilde{d}%
_{L}$ in Eq. (\ref{eqn3.5}), with appropriate changes in all formulas leading
to Eq. (\ref{eqn3.10}). Thus, our SNIa data fitting procedure is valid even if
we use $\widetilde{d}_{L}$ instead of $d_{L}$, which is equivalent to using a
luminosity distance whose estimates are very close to those of standard cosmology.

In KCC, the values of the CG\ parameters $\gamma$ and $\delta$ are also
connected to the current value of the Hubble parameter:%

\begin{align}
H_{0}  &  =\frac{\gamma}{2}c\label{eqn3.14}\\
\mathbf{H}_{0}  &  =\frac{c}{\mathbf{a}_{0}}\delta\nonumber
\end{align}
in SSC or FRW coordinates, respectively, but with $H_{0}\simeq\mathbf{H}_{0}$
for $\left\vert \delta\right\vert \ll1$ \cite{Varieschi:2008va}. Since in this
work we are not relying any longer on the PA data, we can now derive the value
of $\gamma$ directly from the Hubble constant, using the previous equation.

The Union 2.1 SNIa data are consistent with the Hubble constant estimate by
Riess et al. \cite{2011ApJ...730..119R}, $H_{0}=(73.8\pm2.4)\
\operatorname{km}%
\
\operatorname{s}%
^{-1}\ \mathrm{Mpc}^{-1}$, from which we obtain $\gamma=\frac{2}{c}%
H_{0}=\left(  1.596\pm0.052\right)  \times10^{-28}\
\operatorname{cm}%
^{-1}$. However, the most commonly used estimate of the Hubble constant is
from the Planck collaboration 2013 results \cite{Ade:2013zuv}:%

\begin{equation}
H_{0}=(67.3\pm1.2)\
\operatorname{km}%
\
\operatorname{s}%
^{-1}\ \mathrm{Mpc}^{-1}\ \implies\ \gamma=\frac{2}{c}H_{0}=\left(
1.455\pm0.026\right)  \times10^{-28}\
\operatorname{cm}%
^{-1}; \label{eqn3.15}%
\end{equation}
therefore, in the rest of this paper we will consider the value of $\gamma$
above as the current KCC estimate.

It could be argued that, since the Union 2.1 SNIa data are based on the
standard definitions for the luminosity distance, standard candles, etc., it
might be more appropriate to use $d_{REF}=10\ \mathrm{pc}$ as a reference
distance. This leads to a slightly different fitting formula, in view also of
Eqs. (\ref{eqn3.4}) and (\ref{eqn3.14}):%

\begin{align}
\mu(z)  &  =2.5(2+\alpha)\log_{10}(d_{L}/d_{REF})+2.5(1-\beta)\log
_{10}(1+z)\label{eqn3.16}\\
&  =2.5(2+\alpha)\left\{  \log_{10}\left[  \delta\frac{\delta(1+z)+\sqrt
{(1+z)^{2}-(1-\delta^{2})}}{h(1-\delta^{2})}\right]  +8.4768\right\}
\nonumber\\
&  +2.5(1-\beta)\log_{10}(1+z),\nonumber
\end{align}
which also includes the `normalized Hubble constant' $h$ as a fitting
parameter. This dimensionless quantity is related to $H_{0}$ as follows:%

\begin{equation}
H_{0}=100\ h\
\operatorname{km}%
\
\operatorname{s}%
^{-1}\ \mathrm{Mpc}^{-1}=3.2408\times10^{-18}\ h\
\operatorname{s}%
^{-1}. \label{eqn3.17}%
\end{equation}

As in our previous fitting formula (\ref{eqn3.10}), we now have the option of
leaving all four parameters ($\alpha$, $\beta$, $\delta$, and $h$) completely
free, or to fix some of them, for example, by choosing integer values for
$\alpha$ and $\beta$. If we leave all four parameters free, our best fit to
Union 2.1 SNIa data yields:%

\begin{equation}
\alpha=2.005\pm0.253,\ \beta=0.766\pm0.421,\ \delta=3.45\times10^{-5}%
,\ h=0.71, \label{eqn3.18}%
\end{equation}
in line with our previous estimate of the parameters in Eq. (\ref{eqn3.11})
and with our preferred value for $H_{0}$ in Eq. (\ref{eqn3.15}). If we fix the
value of the Hubble constant as in Eq. (\ref{eqn3.15}), i.e., $h=0.673$, and
also set $\alpha=2$,$\ \beta=1$, as it was done in Eq. (\ref{eqn3.13}), we
obtain instead:%

\begin{equation}
\alpha=2,\ \beta=1,\ \delta=\left(  3.367\pm0.008\right)  \times
10^{-5},\ h=0.673. \label{eqn3.19}%
\end{equation}

Comparing our results for $\delta$, in Eqs. (\ref{eqn3.13}) and (\ref{eqn3.19}%
), we see that our two possible fitting formulas (\ref{eqn3.10}) and
(\ref{eqn3.16}) produce consistent results for $\delta\simeq3.36-3.37\times
10^{-5}$, in line also with our previous determinations from Ref.
\cite{Varieschi:2008va}, or in Eq. (\ref{eqn2.15}). In addition, our analysis
confirms that the $f(1+z)$ function in Eq. (\ref{eqn3.8}) should be considered
with $\alpha=2$ and$\ \beta=1$, i.e.,%

\begin{equation}
f(1+z)=\frac{1+z}{\left(  \frac{d_{L}}{d_{REF}}\right)  ^{2}}=\left[
\frac{2\delta}{\delta(1+z)+\sqrt{(1+z)^{2}-(1-\delta^{2})}}\right]  ^{2}(1+z).
\label{eqn3.20}%
\end{equation}

Although our two fitting formulas, Eq. (\ref{eqn3.10}) and Eq. (\ref{eqn3.16}%
), both yield similar results, we have to choose one of the two methods for a
final determination of the CG parameters. Since the former fitting formula
assumes $d_{REF}=\mathbf{a}_{0}\frac{2\delta}{1-\delta^{2}}$, which is more
consistent with the KCC model, while the latter formula assumes $d_{REF}%
=10\ \mathrm{pc}$, which is more consistent with standard cosmology, our final
choice will be the first expression (as it was also done previously in Eq.
(43) of Ref. \cite{Varieschi:2008va}).%

\begin{figure}[ptb]%
\centering
\ifcase\msipdfoutput
\includegraphics[
width=\textwidth
]
{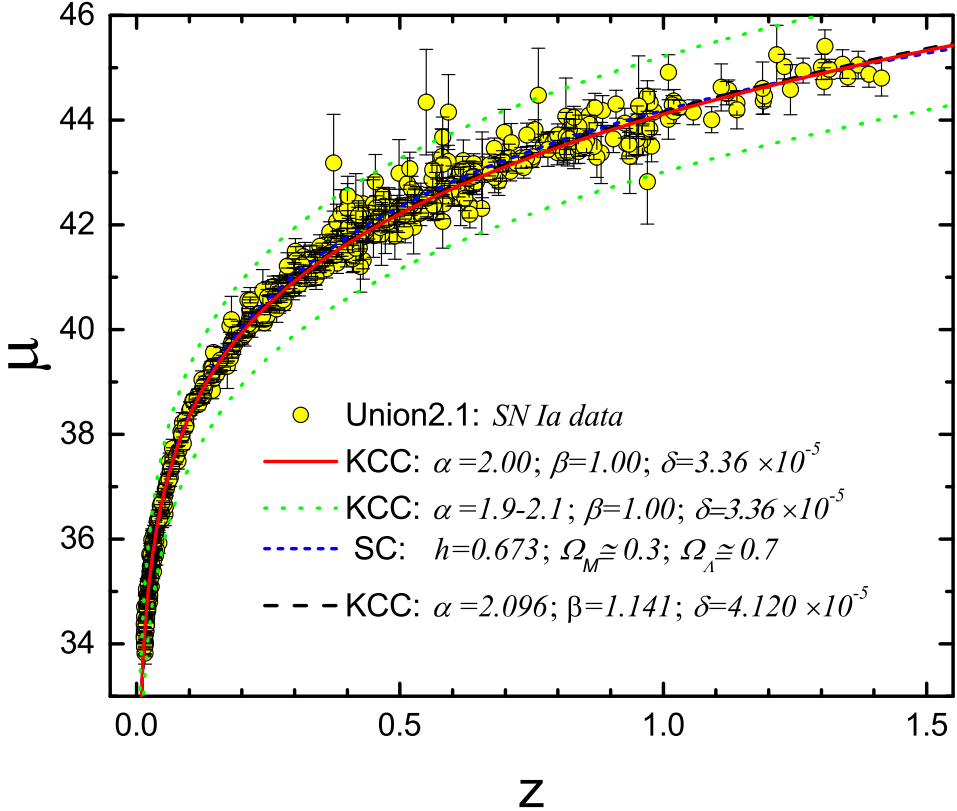}%
\else
\includegraphics[
natheight=4.951800in,
natwidth=6.451700in,
height=4.9518in,
width=6.4517in
]%
{C:/swp55/Docs/KINEMATICAL6/arXiv2/graphics/figure1__1.pdf}%
\fi
\caption{Data from Union 2.1 SNIa set \cite{Suzuki:2011hu} are fitted with Eq.
(\ref{eqn3.10}). Our KCC fits (red-solid for fixed $\alpha$ and $\beta$; black
long-dashed for variable $\alpha$ and $\beta$) show very good statistical
quality ($R^{2}=0.996$) and are very close to the standard cosmology
prediction (SC, blue short-dashed). Also shown (dotted-green curves) is the
range of our KCC fitting curves for a variable $\alpha=1.9-2.1$.}%
\label{fig1}%
\end{figure}

Therefore, in view of Eqs. (\ref{eqn2.12}), (\ref{eqn2.13}), (\ref{eqn3.13}),
and (\ref{eqn3.15}) our revised set of KCC parameters is the following:%

\begin{equation}
\delta=3.36\times10^{-5},\ \gamma=1.46\times10^{-28}%
\operatorname{cm}%
^{-1},\ k=-4.70\times10^{-48}%
\operatorname{cm}%
^{-2},\ \kappa=4.70\times10^{-48}%
\operatorname{cm}%
^{-2}, \label{eqn3.21}%
\end{equation}
and the function $f(1+z)$ is given in Eq. (\ref{eqn3.20}). In the next section
we will plot our results and compare them with those of standard cosmology.

\subsection{\label{sect:plots}Union 2.1 data and KCC plots}

As already mentioned at the beginning of Sect. \ref{sect:Union}, our new
KCC\ fits were performed with the latest Union 2.1 SNIa data\footnote{Also
available in electronic form at: http://supernova.lbl.gov/Union/.}
(\cite{Kowalski:2008ez}, \cite{2010ApJ...716..712A}, \cite{Suzuki:2011hu}).
The Supernova Cosmology Project \textquotedblleft Union2.1\textquotedblright%
\ SNIa compilation is an update of the previous \textquotedblleft
Union2\textquotedblright\ compilation, bringing together data for 833
supernovae, drawn from 19 datasets. Of these, 580 SNe pass usability cuts and
are included in the data set. In Fig. 1 we plot these 580 data points
(distance modulus $\mu$ vs. redshift $z$) together with the standard cosmology
(SC) Hubble plot (blue, short-dashed curve), obtained with standard values of
the critical densities ($\Omega_{M}\cong0.3$, $\Omega_{\Lambda}\cong0.7$) and
with the Hubble constant value in Eq. (\ref{eqn3.15}), i.e., $h=0.673$.%

\begin{figure}[ptb]%
\centering
\ifcase\msipdfoutput
\includegraphics[
width=\textwidth
]
{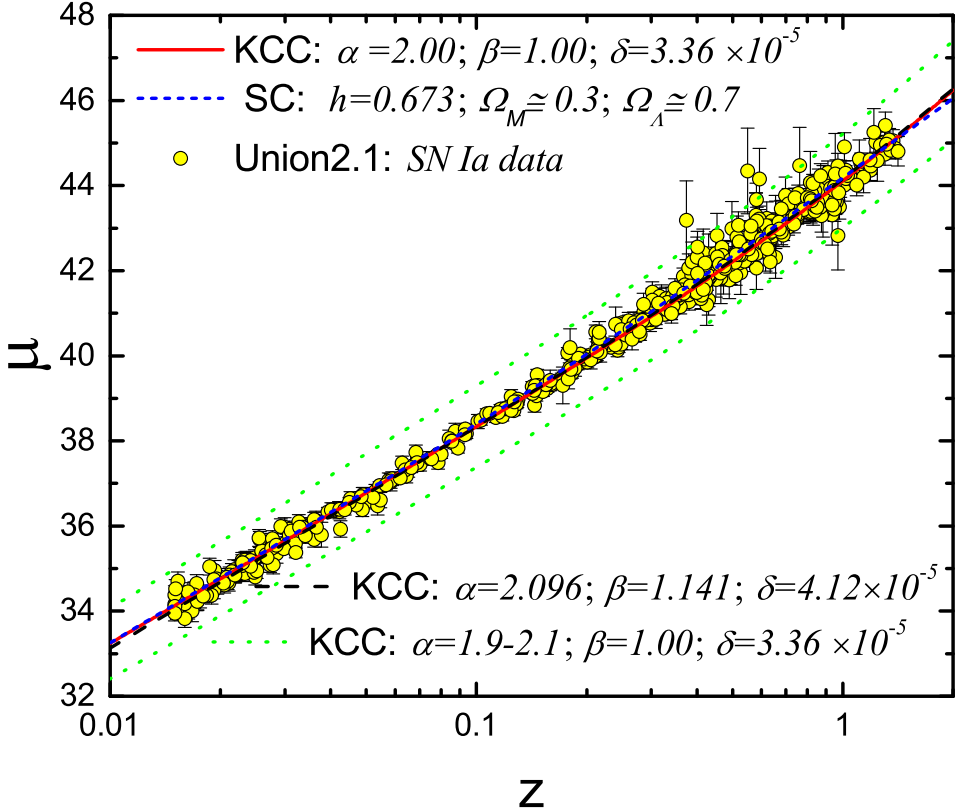}%
\else
\includegraphics[
natheight=4.957300in,
natwidth=6.451700in,
height=4.9573in,
width=6.4517in
]%
{C:/swp55/Docs/KINEMATICAL6/arXiv2/graphics/figure2__2.pdf}%
\fi
\caption{The same data and fitting curves presented in Fig. 1 are shown here
in a standard Hubble plot, with logarithmic axis for the redshift $z$. The
meaning of the symbols and of the different plots is the same as in the
previous figure.}%
\label{fig2}%
\end{figure}

Our KCC fits are also presented in this figure: in red, solid curve, we show
our main fit, using Eq. (\ref{eqn3.10}) and with the values of the parameters
as in Eq. (\ref{eqn3.13}); the green-dotted curves show how our fits depend on
changes of the $\alpha$ parameter (in the range $\alpha=1.9-2.1$), keeping the
other parameters unchanged. Finally, the black, long-dashed curve is our KCC
fit with the parameters as in Eq. (\ref{eqn3.11}), i.e., when all the
parameters are left free in the fitting procedure. This curve is practically
the same as our main KCC fit in solid-red, and both KCC curves are very close
to the standard cosmology theoretical prediction.%

\begin{figure}[ptb]%
\centering
\ifcase\msipdfoutput
\includegraphics[
width=\textwidth
]
{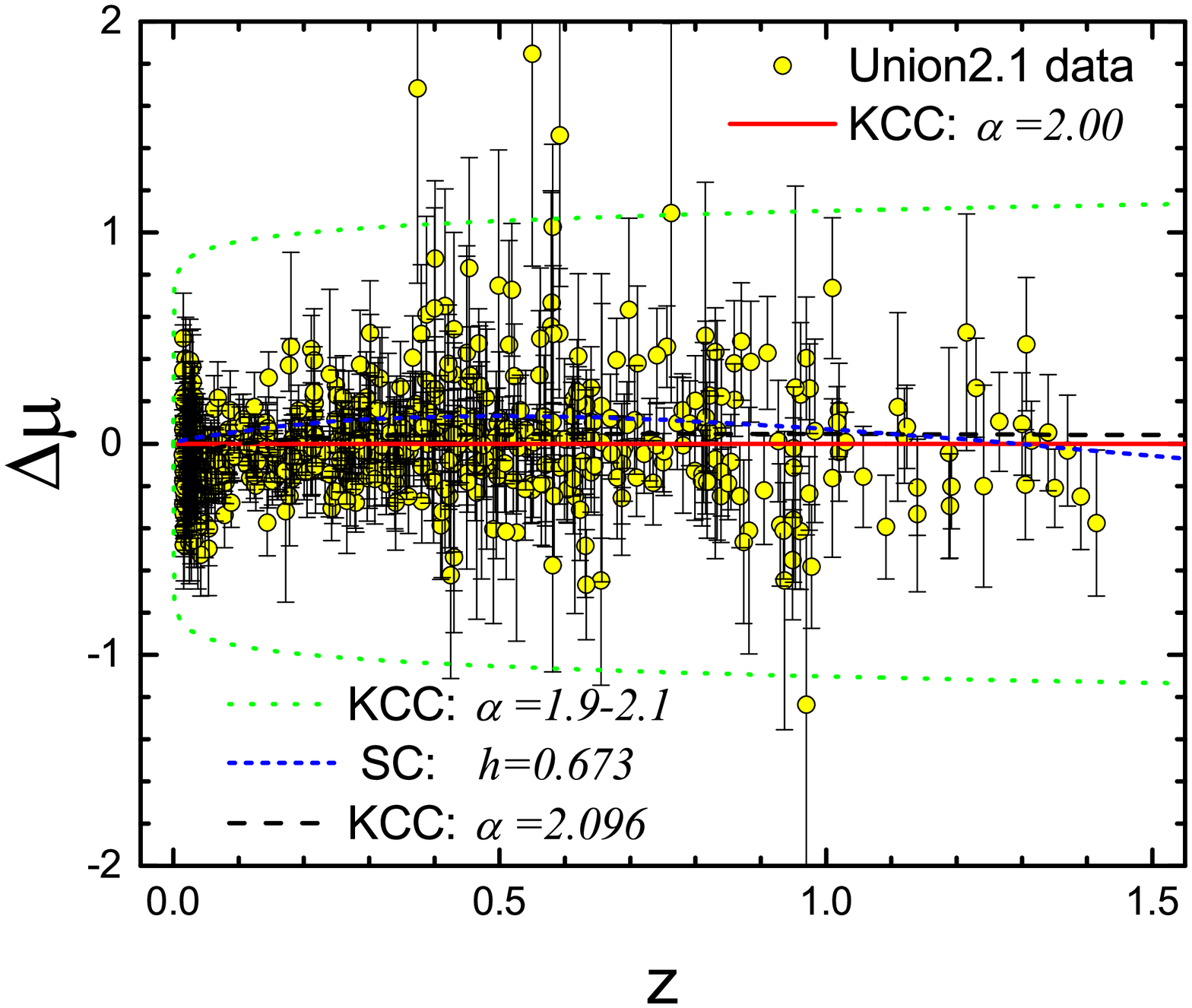}%
\else
\includegraphics[
natheight=4.957300in,
natwidth=6.451700in,
height=4.9573in,
width=6.4517in
]%
{C:/swp55/Docs/KINEMATICAL6/arXiv2/graphics/figure3__3.pdf}%
\fi
\caption{Data from Union 2.1 SNIa set \cite{Suzuki:2011hu} are fitted with Eq.
(\ref{eqn3.10}) and shown as residuals $\Delta\mu$. The baseline is
represented by our main KCC fit (red-solid curve, with parameters as in Eq.
(\ref{eqn3.13})). The meaning of the other curves and symbols is the same as
in the previous figures.}%
\label{fig3}%
\end{figure}

In Fig. 2 we reproduce the same data and the same fitting curves as in Fig. 1,
but in the form of a standard Hubble plot, with logarithmic axis for redshift
$z$. In this way, all the fitting curves become almost straight lines and the
differences between them can be better appreciated. Again, the two main KCC
fits (red-solid and black-long dashed) are almost indistinguishable and only
slightly different from the equivalent standard cosmology prediction (blue, short-dashed).

Similarly, Fig. 3 presents the same information in the form of residual values
$\Delta\mu$, with the baseline represented by our main KCC fit (red-solid,
with parameters as in Eq. (\ref{eqn3.13})). In this figure it is easier to
notice the small differences between our two KCC fits and the standard
cosmology prediction. It is also evident that most of the SNIa data points
fall within the $\alpha=1.9-2.1$ band.

The last study we performed, in connection with the Union 2.1 data, was
related to the low-z behavior of our fitting formulas. As already discussed at
length in \cite{Varieschi:2008va}, we cannot effectively expand in powers of
$z$ our luminosity distance $d_{L}$ in Eq. (\ref{eqn3.4}), due to the very
small value of the $\delta$ parameter. Therefore, we just discard terms
containing $\delta$ in the same expression for $d_{L}$ and retain only the
leading term depending on $z$:%

\begin{equation}
d_{L}\simeq\mathbf{a}_{0}\sqrt{2z}. \label{eqn3.22}%
\end{equation}
Using this expression and $d_{REF}\simeq\mathbf{a}_{0}2\delta$, from Eq.
(\ref{eqn3.7}), in Eq. (\ref{eqn3.10}) and also assuming $\beta=1$, as
suggested by previous fits, we have:%

\begin{equation}
\mu(z)=2.5(2+\alpha)\log_{10}(d_{L}/d_{REF})\simeq2.5(2+\alpha)\log
_{10}\left(  \frac{\sqrt{2z}}{2\delta}\right)  , \label{eqn3.23}%
\end{equation}
which becomes our \textquotedblleft low-z\textquotedblright\ fitting formula.%

\begin{figure}[ptb]%
\centering
\ifcase\msipdfoutput
\includegraphics[
width=\textwidth
]
{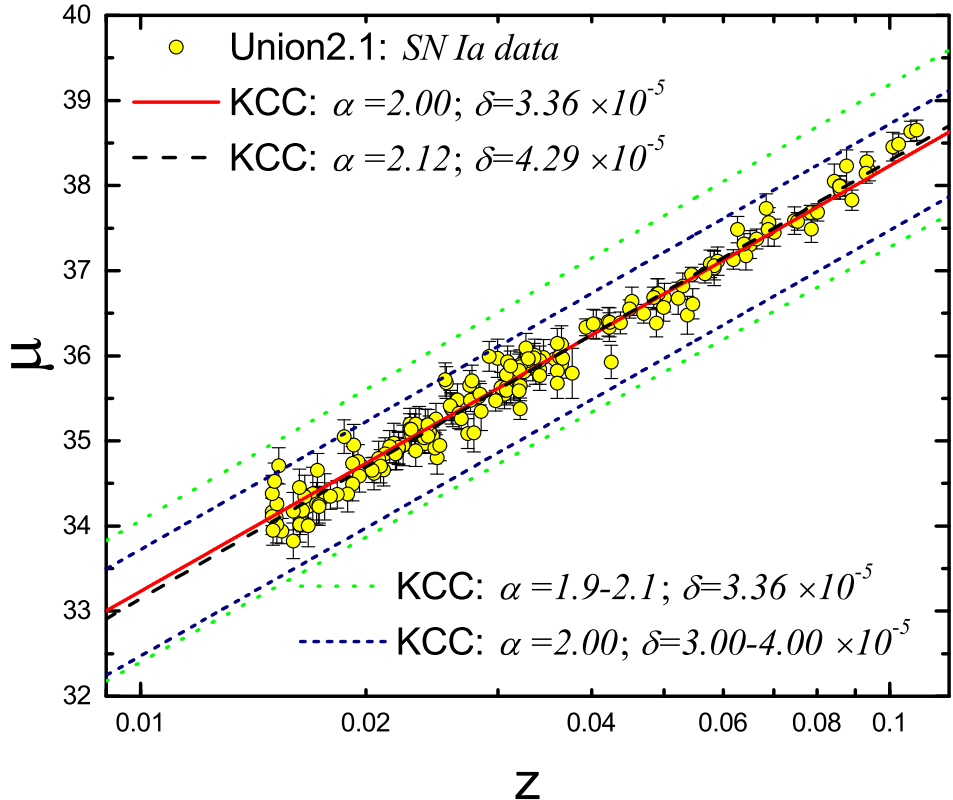}%
\else
\includegraphics[
natheight=4.957300in,
natwidth=6.451700in,
height=4.9573in,
width=6.4517in
]%
{C:/swp55/Docs/KINEMATICAL6/arXiv2/graphics/figure4__4.pdf}%
\fi
\caption{A subset of Union 2.1 SNIa data \cite{Suzuki:2011hu}, for
$z\lesssim0.1$, is fitted with Eq. (\ref{eqn3.23}). Our low-z KCC fits
(red-solid curves for fixed $\alpha$; black long-dashed curves for variable
$\alpha$) yield essentially the same results as in the previous fits, which
used the full range of values for $z$. Also shown are the ranges of our KCC
fitting curves, for a variable $\alpha=1.9-2.1$ (green-dotted curves) and for
a variable $\delta=3.00-4.00\times10^{-5}$ (blue, short-dashed curves).}%
\label{fig4}%
\end{figure}

To check this expression we selected 179 SNIa data from the Union 2.1 set with
$z\lesssim0.1$ and applied our fitting formula (\ref{eqn3.23}) to this data
subset. Fig. 4 shows the results of this low-z fitting: our main KCC fit
(red-solid curve), for a fixed $\alpha=2$, yields $\delta=\left(
3.359\pm0.001\right)  \times10^{-5}$, essentially the same result as in Eq.
(\ref{eqn3.13}) for the whole set of 580 supernovae. Leaving both parameters
free (black, long-dashed curve) yields instead $\alpha=2.121\pm0.040$%
,$\ \delta=\left(  4.287\pm0.340\right)  \times10^{-5}$, and the two KCC
curves almost coincide. In this figure we also show how our low-z fit is
sensitive to the value of $\alpha$ in the range $1.9-2.1$ (green-dotted
curves) and to the value of $\delta$ in the range $3.00-4.00\times10^{-5}$
(blue, short-dashed curves).

In our previous work (see Sect. 3.2 in \cite{Varieschi:2008va}) we also
remarked that our low-z distance modulus expression in Eq. (\ref{eqn3.23}),
for $\alpha=2$, can be rewritten as $\mu(z)\simeq10\log_{10}\left(
\frac{\sqrt{2z}}{2\delta}\right)  =5\log_{10}\left(  \frac{z}{2\delta^{2}%
}\right)  $, so that it corresponds perfectly to the first terms of the
standard cosmology expansion $\mu(z)\simeq25+5\log_{10}\left(  \frac
{cz}{\mathbf{H}_{0}}\right)  =5\log_{10}\left(  10^{5}\frac{cz}{\mathbf{H}%
_{0}}\right)  $, neglecting higher-order terms in $z$. Comparing the
right-hand sides of these two `low-z' expressions, we find a direct connection
between the Hubble constant and the KCC $\delta$ parameter:%

\begin{equation}
\mathbf{H}_{0}\simeq H_{0}=2\times10^{5}c\delta^{2}=67.53\
\operatorname{km}%
\
\operatorname{s}%
^{-1}\ \mathrm{Mpc}^{-1}, \label{eqn3.24}%
\end{equation}
having used our best estimate for $\delta$ in Eq. (\ref{eqn3.13}) and with the
speed of light given as $c=299792.458\
\operatorname{km}%
\operatorname{s}%
^{-1}$.

It is very remarkable that our KCC model and the related SNIa data fitting are
able to obtain an estimate for the Hubble constant which is very close to the
2013 Planck collaboration value. We want to emphasize that our value for
$\delta$ in Eq. (\ref{eqn3.13}) came from the fitting formula in Eq.
(\ref{eqn3.10}), which is independent of any assumed value for $\mathbf{H}_{0}
$.

Therefore, our value of $\mathbf{H}_{0}$ in Eq. (\ref{eqn3.24}) represents
KCC's direct evaluation of the Hubble constant, in agreement with current best
estimates. We can recompute the value for $\gamma$ using $H_{0}=67.53\
\operatorname{km}%
\
\operatorname{s}%
^{-1}\ \mathrm{Mpc}^{-1}$ as $\gamma=\frac{2}{c}H_{0}=1.460\times10^{-28}\
\operatorname{cm}%
^{-1}$, which is essentially equivalent to our previous estimate in Eq.
(\ref{eqn3.15}), based on the 2013 Planck collaboration value for
$\mathbf{H}_{0}$. Following these two estimates, our final value for $\gamma$
will be quoted as $\gamma\simeq1.46\times10^{-28}\
\operatorname{cm}%
^{-1}$, as already reported in Eq. (\ref{eqn3.21}).

\section{\label{sect:hubble}KCC and Hubble parameter data}

Another important test of our KCC model can be performed in relation with
observed data for the Hubble parameter $H(z)$, measured as a function of
redshift. As it was done by Yang et al. in their recent analysis
\cite{Yang:2013skk} of Mannheim's CG, we will use here all the available data
for $H(z)$, obtained from different sources and with different methods, as
reported in Table \ref{TableKey:table1}.\footnote{When both statistical and
systematic errors were quoted (as in \cite{2012JCAP...08..006M},
\cite{2009MNRAS.399.1663G}), we summed these errors in quadrature and reported
the total error in the table.}

Although different methods were used to obtain the data in this table, the
most common argument relies on the fact that the Hubble parameter depends on
the differential age of the Universe, as a function of redshift, in the form:%

\begin{equation}
\mathbf{H}(z)=-\frac{1}{1+z}\frac{dz}{d\mathbf{t}}. \label{eqn4.1}%
\end{equation}%
\begin{table}[tbp] \centering
\begin{tabular}
[c]{||c|c|c|c||}\hline\hline
$z$ & $H(z)\ (%
\operatorname{km}%
\
\operatorname{s}%
^{-1}\ \mathrm{Mpc}^{-1})$ & \textit{Source} & \textit{Method (see
text)}\\\hline
$0.0900$ & $69\pm12$ & Jimenez et al. (2003) \cite{2003ApJ...593..622J} &
DA\\\hline
$0.1700$ & $83\pm8$ & Simon et al. (2005) \cite{2005PhRvD..71l3001S} &
DA\\\hline
$0.2700$ & $77\pm14$ & Simon et al. (2005) \cite{2005PhRvD..71l3001S} &
DA\\\hline
$0.4000$ & $95\pm17$ & Simon et al. (2005) \cite{2005PhRvD..71l3001S} &
DA\\\hline
$0.9000$ & $117\pm23$ & Simon et al. (2005) \cite{2005PhRvD..71l3001S} &
DA\\\hline
$1.3000$ & $168\pm17$ & Simon et al. (2005) \cite{2005PhRvD..71l3001S} &
DA\\\hline
$1.4300$ & $177\pm18$ & Simon et al. (2005) \cite{2005PhRvD..71l3001S} &
DA\\\hline
$1.5300$ & $140\pm14$ & Simon et al. (2005) \cite{2005PhRvD..71l3001S} &
DA\\\hline
$1.7500$ & $202\pm40$ & Simon et al. (2005) \cite{2005PhRvD..71l3001S} &
DA\\\hline
$0.4800$ & $97\pm62$ & Stern et al. (2010) \cite{2010JCAP...02..008S} &
DA\\\hline
$0.8800$ & $90\pm40$ & Stern et al. (2010) \cite{2010JCAP...02..008S} &
DA\\\hline
$0.1791$ & $75\pm4$ & Moresco et al. (2012) \cite{2012JCAP...08..006M} &
DA\\\hline
$0.1993$ & $75\pm5$ & Moresco et al. (2012) \cite{2012JCAP...08..006M} &
DA\\\hline
$0.3519$ & $83\pm14$ & Moresco et al. (2012) \cite{2012JCAP...08..006M} &
DA\\\hline
$0.5929$ & $104\pm13$ & Moresco et al. (2012) \cite{2012JCAP...08..006M} &
DA\\\hline
$0.6797$ & $92\pm8$ & Moresco et al. (2012) \cite{2012JCAP...08..006M} &
DA\\\hline
$0.7812$ & $105\pm12$ & Moresco et al. (2012) \cite{2012JCAP...08..006M} &
DA\\\hline
$0.8754$ & $125\pm17$ & Moresco et al. (2012) \cite{2012JCAP...08..006M} &
DA\\\hline
$1.0370$ & $154\pm20$ & Moresco et al. (2012) \cite{2012JCAP...08..006M} &
DA\\\hline
$0.2400$ & $79.69\pm2.65$ & Gazta$\tilde{n}$aga et al. (2009)
\cite{2009MNRAS.399.1663G} & BAO\\\hline
$0.4300$ & $86.45\pm3.68$ & Gazta$\tilde{n}$aga et al. (2009)
\cite{2009MNRAS.399.1663G} & BAO\\\hline
$0.0700$ & $69\pm19.6$ & Zhang et al. (2012) \cite{Zhang:2012mp} & DA\\\hline
$0.1200$ & $68.6\pm26.2$ & Zhang et al. (2012) \cite{Zhang:2012mp} &
DA\\\hline
$0.2000$ & $72.9\pm29.6$ & Zhang et al. (2012) \cite{Zhang:2012mp} &
DA\\\hline
$0.2800$ & $88.8\pm36.6$ & Zhang et al. (2012) \cite{Zhang:2012mp} &
DA\\\hline
$0.4400$ & $82.6\pm7.8$ & Blake et al. (2012) \cite{2012MNRAS.425..405B} & BAO
and GC\\\hline
$0.6000$ & $87.9\pm6.1$ & Blake et al. (2012) \cite{2012MNRAS.425..405B} & BAO
and GC\\\hline
$0.7300$ & $97.3\pm7.0$ & Blake et al. (2012) \cite{2012MNRAS.425..405B} & BAO
and GC\\\hline
$0.3500$ & $82.1\pm5$ & Chuang et al. (2012) \cite{Chuang:2011fy} &
GC\\\hline\hline
\end{tabular}
\caption{Available Hubble parameter data H(z), from various sources,
obtained with different methods.}\label{TableKey:table1}%
\end{table}%

Therefore, a determination of $\frac{dz}{d\mathbf{t}}$, or more practically of
the ratio $\frac{\Delta z}{\Delta\mathbf{t}}$ between finite intervals of
redshift and time, will lead to a direct measurement of $\mathbf{H}(z)$.

In order to measure the time interval $\Delta\mathbf{t}$, we need to identify
and use so-called `cosmic chronometers,' i.e., astrophysical objects, such as
a galaxies, whose evolution follows a known fiducial model, so that these
objects behave as `standard clocks' in the Universe.

Once this population of standard clocks has been found and dated, the
`differential-age' technique can be used: the age difference $\Delta
\mathbf{t}$, and the corresponding redshift difference $\Delta z$, between two
of these cosmic chronometers can be measured, thus determining $\mathbf{H}(z)$
in view of Eq. (\ref{eqn4.1}). This differential age (DA) method has the
advantage of not using any integrated cosmological quantity (such as the
luminosity distance, which is expressed through an integral in standard
cosmology), since these quantities depend on the integral of the expansion
history, thus yielding less direct measurements of the expansion history itself.

Since the original proposal of this DA method (\cite{Jimenez:2001gg},
\cite{2003ApJ...593..622J}), the best choice of `cosmic chronometers' was
found to be a population of `red-envelope' galaxies: massive galaxies,
harbored in high-density regions of galaxy clusters and containing the oldest
stellar populations, which are now evolving only passively (i.e., with very
limited new star formation). The age of these passively evolving galaxies can
then be used in connection with the DA technique explained above to measure
$\mathbf{H}(z)$ (\cite{2003ApJ...593..622J}, \cite{2005PhRvD..71l3001S},
\cite{2010JCAP...02..008S}). A similar approach, also based on passively
evolving galaxies, but more centered on a differential spectroscopic evolution
of early-type galaxies as a function of redshift, was introduced by Moresco
et. al. (\cite{Moresco:2010wh}, \cite{2012JCAP...08..006M}), yielding more
data points, followed by the more recent work by Zhang et al.
\cite{Zhang:2012mp}.

A different approach \cite{2009MNRAS.399.1663G} to the measurement of
$\mathbf{H}(z)$ considered instead the baryon acoustic oscillations (BAO) peak
position as a standard ruler in the radial direction. This BAO method was
later connected to the Alcock-Paczynski distortion from galaxy clustering (GC)
in the WiggleZ Dark Energy Survey \cite{2012MNRAS.425..405B}, and one
additional data point was recently obtained \cite{Chuang:2011fy} by using
galaxy clustering data. All the measured data points for $\mathbf{H}(z)$ are
reported in Table \ref{TableKey:table1}; we will now interpret these data in
view of our kinematical conformal cosmology.

In KCC, the Hubble parameter is directly related to $z$ as follows (see
Eq.(10) in \cite{Varieschi:2008va}):%

\begin{equation}
\mathbf{H}(z)=\frac{c}{\mathbf{a}_{0}}\sqrt{(1+z)^{2}-(1-\delta^{2})}%
=\frac{\mathbf{H}_{0}}{\delta}\sqrt{(1+z)^{2}-(1-\delta^{2})}, \label{eqn4.2}%
\end{equation}
in view also of Eq. (\ref{eqn3.14}) and assuming $\delta>0$. At first, it
seems impossible to fit the observational Hubble data (OHD) in Table
\ref{TableKey:table1} with the formula on the right-hand side of the last
equation, for $\delta\sim10^{-5}$ and $\mathbf{H}_{0}$ close to standard
values. However, the OHD are obtained essentially from Eq. (\ref{eqn4.1}), or
rather from the critical determination of the time interval $\Delta
\mathbf{t}\approx d\mathbf{t}$, which enters the denominator on the right-hand
side of this equation.

Although the differential age methods used to obtain these OHD in the
literature are slightly different (and even more different are the methods
based on BAO and/or GC), they all rely heavily on time, distance, and
spectroscopic determinations, based on standard cosmology. Since KCC allows
for intrinsic scaling of lengths, time intervals, energies, luminosities,
etc., as in Eqs. (\ref{eqn3.1})-(\ref{eqn3.3}), we need to allow the presence
of these scaling factors, such as powers of $(1+z)$ and/or $f(1+z)$, into our
fitting formula (\ref{eqn4.2}).

In view also of the general form of $f(1+z)$ in Eq. (\ref{eqn3.8}), we
generalize our fitting formula for $\mathbf{H}(z)$\ as:%

\begin{align}
\mathbf{H}(z)  &  =\frac{\mathbf{H}_{0}}{\delta}\frac{\left(  1+z\right)
^{l}}{\left(  \frac{d_{L}}{d_{REF}}\right)  ^{m}}\sqrt{(1+z)^{2}-(1-\delta
^{2})}\label{eqn4.3}\\
&  =2\times10^{5}c\delta(1+z)^{l}\left[  \frac{2\delta}{\delta(1+z)+\sqrt
{(1+z)^{2}-(1-\delta^{2})}}\right]  ^{m}\sqrt{(1+z)^{2}-(1-\delta^{2}%
)},\nonumber
\end{align}
where $l$ and $m$ are free parameters to be determined with our fitting
procedure. The additional factor of $\left(  1+z\right)  ^{l}/\left(
d_{L}/d_{REF}\right)  ^{m}$, introduced in the last equation, is justified in
the same way as it was done in Eq. (\ref{eqn3.8}): it is just a reasonable
`ansatz' based on the only two functions of $(1+z)$ introduced in KCC. Of
course, the new parameters $l$ and $m$ in Eq. (\ref{eqn4.3}) are not
necessarily related to the similar $\alpha$ and $\beta$ parameters used before
with the SNIa data, since we are now fitting a different type of astrophysical
data. In the last equation we also used our direct connection in Eq.
(\ref{eqn3.24}) between $\mathbf{H}_{0}$ and $\delta$ to avoid
over-parametrizing this fitting formula.%

\begin{figure}[ptb]%
\centering
\ifcase\msipdfoutput
\includegraphics[
width=\textwidth
]
{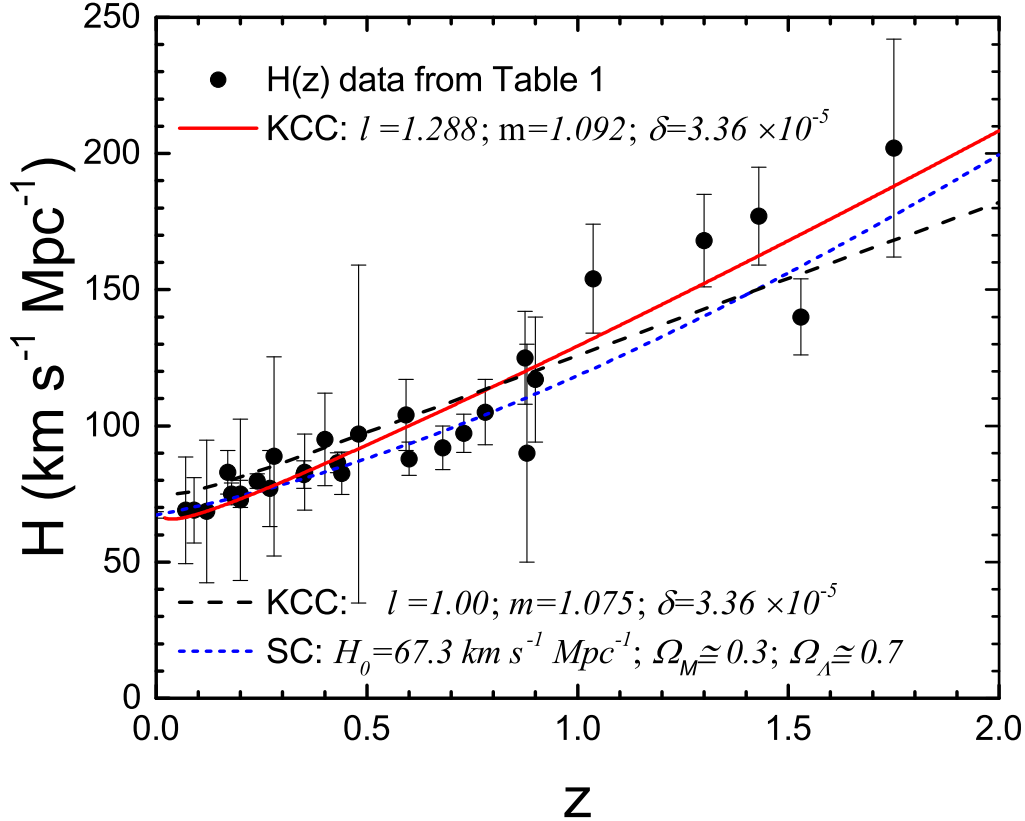}%
\else
\includegraphics[
natheight=4.957300in,
natwidth=6.451700in,
height=4.9573in,
width=6.4517in
]%
{C:/swp55/Docs/KINEMATICAL6/arXiv2/graphics/figure5__5.pdf}%
\fi
\caption{OHD\ from Table \ref{TableKey:table2} are fitted with our KCC\ Eq.
(\ref{eqn4.3}), using values for the parameters as in Eq. (\ref{eqn4.4})
(red-solid curve), or as in Eq. (\ref{eqn4.5}) (black, long-dashed curve).
Also shown is the standard cosmology prediction (blue, short-dashed curve).}%
\label{fig5}%
\end{figure}

We then used our revised formula (\ref{eqn4.3}) to fit the OHD in Table
\ref{TableKey:table1}, allowing up to three dimensionless parameters: $\delta
$, $l$, and $m$. However, leaving all three parameters completely free does
not lead to a satisfactory fit of the data, so we simply set $\delta$ to our
preferred value: $\delta=3.36\times10^{-5}$. Our best fit, considering $l$ and
$m$ as free parameters, is:%

\begin{equation}
l=1.288\pm0.084,\ m=1.092\pm0.006,\ \delta=3.36\times10^{-5}, \label{eqn4.4}%
\end{equation}
and is shown in Fig. 5 (red-solid curve), together with all the OHD from Table
\ref{TableKey:table1}. If we fix $l$ to be an integer value, close to the
previous estimate, we obtain instead:%

\begin{equation}
l=1,\ m=1.075\pm0.003,\ \delta=3.36\times10^{-5}, \label{eqn4.5}%
\end{equation}
which is also shown in Fig. 5 (black, long-dashed curve). In the same figure,
the standard cosmology prediction, $\mathbf{H}(z)=\mathbf{H}_{0}\sqrt
{\Omega_{M}(1+z)^{3}+\Omega_{\Lambda}+\Omega_{R}(1+z)^{4}+\Omega_{K}(1+z)^{2}%
}$, is shown (blue, short-dashed curve) for $\Omega_{M}\cong0.3$,
$\Omega_{\Lambda}\cong0.7$, $\Omega_{R}=\Omega_{K}\approx0$, and
$\mathbf{H}_{0}=67.3\
\operatorname{km}%
\
\operatorname{s}%
^{-1}\ \mathrm{Mpc}^{-1}$.

The $l\simeq1$ value for the first parameter in our KCC fitting formula can be
explained as originating from the scaling law of time intervals, $\Delta
t_{z}=(1+z)\ \Delta t_{0}$, applied to the measured value $\Delta
\mathbf{t}\approx d\mathbf{t}$ which enters Eq. (\ref{eqn4.1}). In other
words, the observed age differences at redshift $z$ are actually $\Delta
t_{z}$ intervals, but the age intervals entering Eq. (\ref{eqn4.1}) should be
considered as $\Delta t_{0}$ intervals, since standard cosmology does not
allow for rescaled quantities. Thus, combining Eqs. (\ref{eqn4.1}) and
(\ref{eqn4.2}), we have: $\mathbf{H}(z)=-\frac{1}{1+z}\frac{\Delta z}%
{\Delta\mathbf{t}_{0}}=(1+z)\frac{\mathbf{H}_{0}}{\delta}\sqrt{(1+z)^{2}%
-(1-\delta^{2})}$, where the $(1+z)$ factor on the right-hand side is due to
the rescaling of the time intervals.

The $m\approx1$ value for the second parameter in our KCC fits is not so
easily explained. This corresponds to a factor $1/\left(  \frac{d_{L}}%
{d_{REF}}\right)  ^{m}$ on the right-hand side of our fitting formula
(\ref{eqn4.3}), with $m$ close to unity. This could be due to the fact that
the OHD are determined through spectroscopic measurements (involving the
scaling factor $f(1+z)=(1+z)/\left(  d_{L}/d_{REF}\right)  ^{2}$), or because
the age determinations of the cosmic chronometers, such as the `red-envelope'
galaxies, involve their luminosity distances, thus allowing for the KCC
correction factor $\left(  d_{L}/d_{REF}\right)  $ to appear in our fitting formula.

In particular, age estimates are typically sensitive to the distance scale
(see discussion in Ref. \cite{Weinberg2}, pp. 62-63):\ a fractional change
$\delta d/d$ in distance estimates will produce a change $\delta L/L=-2\delta
d/d$ in absolute luminosities and thus a fractional change $\delta
t/t\approx+2\delta d/d$ in age estimates, since the absolute luminosity of
stars at the turn-off point in the main sequence is roughly inversely
proportional to the age of the globular cluster being studied. In KCC the
change in luminosity distance $\delta d_{L}$ is due to the difference between
the revised $d_{L}=\sqrt{\frac{f(1+z)}{(1+z)}\frac{L_{0}}{4\pi l}}$ of Eq.
(\ref{eqn3.4}) and the standard cosmology expression $d_{L}=\sqrt{\frac{L_{0}%
}{4\pi l}}$, which assumes an invariable luminosity $L_{0}$.

Therefore, in view also of Eq. (\ref{eqn3.20}), a fractional change $\delta
d_{L}/d_{L}=\sqrt{\frac{f(1+z)}{(1+z)}}-1=1/\left(  \frac{d_{L}}{d_{REF}%
}\right)  -1$ might introduce a correcting factor $1+\delta d_{L}%
/d_{L}=1/\left(  \frac{d_{L}}{d_{REF}}\right)  $ into our CG age estimates and
ultimately yield a corresponding factor $1/\left(  \frac{d_{L}}{d_{REF}%
}\right)  ^{m}$ on the right-hand side of our fitting formula (\ref{eqn4.3}),
with $m\approx1$. Due to the complexity of the details related to the
experimental measurements of the OHD, at this point we are unable to further
explain the presence of this factor in our fitting formula for $\mathbf{H}(z)$.

Finally, we wish to comment on the `age problem' analyzed in Ref.
\cite{Yang:2013skk}, which was related to Mannheim's CG. The issue being
studied was a possible age problem for the old quasar APM\ 08279+5255 at
$z=3.91$, as well as the current estimates of the age of the Universe. As
already remarked in Sect. \ref{sect:introduction}, it was shown that CG does
not suffer from an age problem, as it might be the case instead for standard
cosmology (see again \cite{Yang:2013skk}\ and references therein).

For a cosmological model where $\mathbf{H}(z)$ is known explicitly, all age
estimates are essentially obtained by integrating Eq. (\ref{eqn4.1}). For
instance, the current age of the Universe $\mathbf{t}_{0}$ is:%

\begin{equation}
\mathbf{t}_{0}=T(0,\infty)=-\int\limits_{\infty}^{0}\frac{1}{\left(
1+z\right)  \mathbf{H}(z)}dz, \label{eqn4.6}%
\end{equation}
assuming $z=\infty$ at time zero and $z=0$ at current time. More generally,
the age of an astrophysical object (such as the old quasar mentioned above)
which is observed at redshift $z$, but whose formation occurred at earlier
times, corresponding to a formation redshift $z_{f}>z$, is computed as:%

\begin{equation}
T(z,z_{f})=-\int\limits_{z_{f}}^{z}\frac{1}{\left(  1+z\right)  \mathbf{H}%
(z)}dz. \label{eqn4.7}%
\end{equation}

In $\Lambda$CDM cosmology, using the standard expression for $\mathbf{H}%
(z)$\ with $\Omega_{M}\cong0.3$, $\Omega_{\Lambda}\cong0.7$, $\Omega
_{R}=\Omega_{K}\approx0$, and $\mathbf{H}_{0}=67.3\
\operatorname{km}%
\
\operatorname{s}%
^{-1}\ \mathrm{Mpc}^{-1}$, the age of the Universe from Eq. (\ref{eqn4.6}) is
computed as $\mathbf{t}_{0}=14.0\ \mathrm{Gyr}$, in line with estimates based
on globular clusters, or other astrophysical objects. On the contrary, the
quasar APM\ 08279+5255 is observed at $z=3.91$, with an estimated formation
redshift $z_{f}=15$ \cite{Yang:2013skk}. Using Eq. (\ref{eqn4.7}), the
standard cosmology age for this quasar would be $T_{SC}%
(3.91,15)=1.34\ \mathrm{Gyr}$, causing a possible age problem, since the best
estimated age for this quasar is $2.1\ \mathrm{Gyr}$, with a $1\sigma$ lower
limit of $1.8\ \mathrm{Gyr}$ and an absolute lowest limit of
$1.5\ \mathrm{Gyr}$ \cite{Yang:2013skk}.

As discussed at length in our previous work (see Sect. 4.5 in Ref.
\cite{Varieschi:2008fc}), in KCC we have two possible time coordinates: the
static standard coordinate $t$ related to our local unit of time, as opposed
to the FRW coordinate $\mathbf{t}$, where the former is essentially the
conformal time of the latter. When using the former coordinate $t$, the
Universe does not appear to have initial or final singularities (thus, the age
of the Universe would be infinite, if measured using this coordinate), while
both singularities appear when using the latter coordinate $\mathbf{t}$.

However, if we use FRW coordinates to estimate ages, i.e., if we use
$\mathbf{H}(z)$ as in Eq. (\ref{eqn4.2}) in Eqs. (\ref{eqn4.6})-(\ref{eqn4.7}%
), we would obtain extremely small estimates for the age of the Universe and
for the age of the quasar being studied. This shows that age estimates in KCC
are not directly comparable with age estimates in SC, in the same way that
luminosity distances in KCC and SC are widely different, as already mentioned
in Sect. \ref{sect:supernovae}.

Once again, to reconcile the two different views, we must use the `revised'
formula for $\mathbf{H}(z)$ in Eq. (\ref{eqn4.3}) with the KCC parameters
determined in Eq. (\ref{eqn4.4}), or Eq. (\ref{eqn4.5}). Using this formula
and the related parameters in the age equations (\ref{eqn4.6}) and
(\ref{eqn4.7}) yields the results reported in Table \ref{TableKey:table2} (the
corresponding SC\ results are also shown in this table).%

\begin{table}[tbp] \centering
\begin{tabular}
[c]{||c|c|c||}\hline\hline
Model & Age of Universe & Age of Quasar\\\hline
SC ($\Omega_{M}\cong0.3$, $\Omega_{\Lambda}\cong0.7$, $\mathbf{H}_{0}=67.3\
\operatorname{km}%
\
\operatorname{s}%
^{-1}\ \mathrm{Mpc}^{-1}$) & $14.0\ \mathrm{Gyr}$ & $1.34\ \mathrm{Gyr}%
$\\\hline
KCC - parameters from Eq. (\ref{eqn4.4}) & $14.2\ \mathrm{Gyr}$ &
$1.65\ \mathrm{Gyr}$\\\hline
KCC - parameters from Eq. (\ref{eqn4.5}) & $15.8\ \mathrm{Gyr}$ &
$2.45\ \mathrm{Gyr}$\\\hline\hline
\end{tabular}
\caption{Standard Cosmology and KCC estimates for the age of the Universe
and of quasar APM 08279+5255.}\label{TableKey:table2}%
\end{table}%

As it can be seen from the values in this table, the KCC age of the Universe
is, in both cases, in agreement with the accepted estimates. In KCC, there is
also no apparent age problem for the Quasar APM\ 08279+5255: our first
estimate ($1.65\ \mathrm{Gyr}$) is greater than the lowest age limit of
$1.5\ \mathrm{Gyr}$, while our second estimate ($2.45\ \mathrm{Gyr}$) is
larger than the best estimated age for this quasar of $2.1\ \mathrm{Gyr}$.

\section{\label{sect:conclusions}Conclusions}

In this work we analyzed KCC in view of recent astrophysical data from SNIa
and determinations of the Hubble parameter as a function of redshift. The
analysis of the supernova data essentially confirmed our previous work on the
subject, but this time we used the recent Union 2.1 data (580 data points,
instead of 292) and more general assumptions for our KCC fitting formulas.

It was shown that the KCC model can again accommodate all existing SNIa data,
without resorting to dark energy, or to any other exotic component of the
Universe. Moreover, the current value of the Hubble constant was derived
directly from the SNIa data, using the KCC model, without any prior assumption
for this value. We obtained a KCC estimate of the Hubble constant as
$\mathbf{H}_{0}=67.53\
\operatorname{km}%
\
\operatorname{s}%
^{-1}\ \mathrm{Mpc}^{-1}$, very close to the 2013 Planck collaboration value.
The other KCC fundamental parameters, $\delta$, $\gamma$, $\kappa$, and $k$,
were critically re-evaluated and their updated values reported in Eq.
(\ref{eqn3.21}).

KCC was also tested against OHD for $\mathbf{H}(z)$ and in relation with the
age of the Universe and of old quasars. As in the case of luminosity distance
determinations, it was found that age determinations in KCC need to be
corrected by using the same scale factors which are at the basis of our model.
With these scale corrections, KCC can effectively accommodate the existing
$\mathbf{H}(z)$ data, and does not show any apparent age problem, including
the case of quasar APM\ 08279+5255.

Therefore, our final conclusion is that kinematical conformal cosmology is
still a viable alternative cosmological model, although surely not as popular
as other models based on conformal gravity, or standard $\Lambda$CDM
cosmology. Further studies will be needed to check this model against other
astrophysical data in order to see if it remains a possible alternative cosmology.

\acknowledgments{The author would like to thank Loyola Marymount University and the Seaver College of Science and Engineering for continued support and for granting a sabbatical leave of absence to the author, during which this work was completed. The author is indebted to Ms. Z. Burstein for helpful comments and for proofreading the original manuscript. Finally, the author also thanks the anonymous referees for their valuable suggestions and useful comments, which helped improve the final version of this paper.}



\bibliographystyle{apsrev}
\bibliography{CONFORMAL,FLYBY,KINEMATICAL6,MANNHEIM_RECENT,PIONEER,VARIESCHI}

\end{document}